\documentclass[prd,aps,superscriptaddress,showpacs,amssymb,
amsmath,twocolumn,floatfix]{revtex4}
\usepackage{bm,graphicx,dcolumn}

\newcommand{\Di}{\displaystyle}
\newcommand{\la}{\left\langle}
\newcommand{\ra}{\right\rangle}
\newcommand{\matel}[1]{\la 0 \left| #1 \right| \bar{Q}Q \ra}
\newcommand{\qcd}{\text{QCD}}
\newcommand{\nrqcd}{\text{NRQCD}}

\newcommand{\Ivertex}{I_\text{vertex}}
\newcommand{\Iearlobe}{I_\text{earlobe}}
\newcommand{\Ibubble}{I_\text{bubble}}
\newcommand{\Itadpole}{I_\text{tadpole}}

\newcommand{\Iz}{I_\text{Z}}
\newcommand{\Iodd}{I_\text{odd}}
\newcommand{\Iin}{I_\text{in}}
\newcommand{\Iout}{I_\text{out}}
\newcommand{\Iqcd}{I_\qcd}
\newcommand{\Inrqcd}{I_\nrqcd}

\newlength{\myfigwidth}
\begin{document}

\title{Leptonic widths of heavy quarkonia:
S-Wave QCD/NRQCD matching coefficients for the electromagnetic vector 
annihilation current at $\mathcal{O}(\alpha_s v^2)$}

\author{A. \surname{Hart}}
\affiliation{SUPA, School of Physics, University of Edinburgh, Edinburgh EH9
  3JZ, United Kingdom}
\author{G.M. \surname{von Hippel}}
\affiliation{Department of Physics,
  University of Regina, Regina, Saskatchewan S4S 0A2, Canada}
\author{R.R. \surname{Horgan}}
\affiliation{Department of Applied Mathematics and Theoretical Physics,
University of Cambridge, Centre for Mathematical Sciences, Cambridge
CB3 0WA, United Kingdom}

\preprint{DAMTP-2006-...}
\preprint{Edinburgh 2006/10}

\begin{abstract}
  We construct the S-wave part of the electromagnetic vector
  annihilation current to $\mathcal{O}(\alpha_s v^2)$ on the lattice
  for heavy quarks whose dynamics are described by the NRQCD action,
  and $v$ is the non-relativistic quark velocity inside the meson. The
  lattice vector current for $Q{\bar Q}$ annihilation is expressed as
  a linear combination of lattice operators with quantum numbers $L=0,
  J^P=1^-$, and the coefficients are determined by matching this
  lattice current to the corresponding continuum current in QCD to
  $O(v^2)$ at one-loop. The annihilation channel gives a complex
  amplitude and a proper choice for the contours of integration is
  needed; a simple Wick rotation is not possible. In this way, and
  with a careful choice of subtraction functions in the numerical
  integration, the Coulomb-exchange and infrared singularities
  appearing in the amplitudes are successfully treated. The matching
  coefficients are given as a function of the heavy quark mass $Ma$ in
  lattice units.  An automated vertex generation program written in
  \textsc{Python} is employed, allowing us to use a realistic NRQCD
  action and an improved gluon lattice action. A change in the
  definition of either action is easily accommodated in this
  procedure. The final result, when combined with lattice simulation
  results, describes the electromagnetic decays of heavy quarkonia,
  notably the $\Upsilon$ meson.
\end{abstract}

\maketitle


\newcommand{\BE}{\begin{equation}}
\newcommand{\EE}{\end{equation}}
\newcommand{\be}{\begin{equation}}
\newcommand{\ee}{\end{equation}}

\newcommand{\BEA}{\begin{eqnarray}}
\newcommand{\EEA}{\end{eqnarray}}
\newcommand{\bea}{\begin{eqnarray}}
\newcommand{\eea}{\end{eqnarray}}

\newcommand{\Tr}{\mathrm{Tr}}
\newcommand{\tr}{\mathrm{tr}}
\newcommand{\Det}{\mathrm{Det}}

\newcommand{\ad}{\mathbf{ad}}

\renewcommand{\Re}{\mathrm{Re}}
\renewcommand{\Im}{\mathrm{Im}}

\newcommand{\mint}[1]{\int\!\!\frac{d^3#1}{(2\pi)^3}}
\newcommand{\fmint}[1]{\int\!\!\frac{d^4#1}{(2\pi)^4}}
\newcommand{\dmint}[1]{\int\!\!\frac{d^d#1}{(2\pi)^d}}

\newcommand{\Fint}[1]{\int\!\!\mathcal{D}#1}

\newcommand{\mb}[1]{\mathbf{#1}}
\newcommand{\ds}[1]{\mathds{#1}}
\newcommand{\mc}[1]{\mathcal{#1}}

\newcommand{\C}{\mathcal{C}}
\newcommand{\D}{\mathcal{D}}
\newcommand{\F}{\mathcal{F}}
\renewcommand{\L}{\mathcal{L}}
\newcommand{\M}{\mathcal{M}}
\renewcommand{\O}{\mathcal{O}}

\def\slashchar#1{\setbox0=\hbox{$#1$}           
   \dimen0=\wd0                                 
   \setbox1=\hbox{/} \dimen1=\wd1               
   \ifdim\dimen0>\dimen1                        
      \rlap{\hbox to \dimen0{\hfil/\hfil}}      
      #1                                        
   \else                                        
      \rlap{\hbox to \dimen1{\hfil$#1$\hfil}}   
      /                                         
   \fi}                                         %

\section{Introduction}

Heavy quark states like the $J/\Psi$ \cite{aubert:jpsi,augustin:jpsi}
and $\Upsilon$ \cite{carlson:upsilon,herb:upsilon} mesons play a
central role in the experimental study of the electroweak
interactions. It is therefore important that we have reliable
non-perturbative QCD predictions of their properties against which to
compare. Lattice Monte Carlo simulations provide the only
systematically improvable framework for such studies, but relativistic
quark actions do not lend themselves very easily to lattice
simulations of heavy quark dynamics; the Compton wavelengths of heavy
quarks are small compared to currently feasible lattice spacings.

Fortunately the heavy quarks are much heavier than the hadronic scale
$\Lambda\approx 200$~MeV, whilst their kinetic energy is small (as
demonstrated by the radial excitations of the mesons being much
smaller than the ground state energy). This allows a non-relativistic
description of the mesons using the NRQCD effective field theory
\cite{Caswell:1985ui,Lepage:1992tx}, 
using the heavy quark velocity as the expansion parameter.

Simply put, the goal of this paper is to provide matching coefficients
that allow NRQCD matrix elements (calculated non-perturbatively in a
lattice simulation) to be used to predict heavy quark phenomenology,
in particular the leptonic widths of the $\Upsilon$ mesons.

More precisely, to obtain accurate results from a lattice simulation
the QCD and NRQCD actions must be systematically improved to eliminate
errors due to lattice artifacts, relativistic corrections and
radiative effects.  Both perturbative and non-perturbative methods
exist to do this.  A similar programme is needed for improvement of
lattice operators and currents. In this paper we use perturbation
theory to match matrix elements of the S-wave part of the vector
$Q\bar{Q}$ heavy-heavy electromagnetic annihilation current calculated
on the lattice to the continuum result, ensuring that the lattice
results give the correct answer to $\mathcal{O}(\alpha_s)$ in the
strong coupling constant and $\mathcal{O}(v^2)$ in the velocity. This
technique has already been used to improve the weak annihilation
current for leptonic B-meson decay
\cite{Morningstar:1997ep,Morningstar:1998yx} 
via the weak annihilation of a heavy quark $Q$ and light anti-quark
$\bar{q}$.

The $Q\bar{Q}$ annihilation is more complicated than the weak
$Q\bar{q}$ case. In the heavy-light case, we can exploit the crossing
symmetry of the relativistic light quark action to match instead the
weak heavy-light $Qq$ form-factor. The amplitude for this is purely
real and so the choice of integration contour for the temporal
component of the momentum is straightforward (parallel to the
imaginary axis).  The NRQCD action lacks this crossing symmetry and so
the time-like improved lattice vector current (relevant for
annihilation) is not \textit{a priori} related to its space-like
counterpart (which determines the heavy quark form-factor). The
amplitude for on-shell $Q\bar{Q}$ annihilation is complex with a
threshold for $Q\bar{Q}$ scattering and has the additional
complication that it contains a Coulomb singularity. The calculation
therefore entails a careful choice for the integration contours. For
heavy quark velocities $v>0$ this does not correspond to the simple
Wick rotation (generally with constant real part displacement) which
suffices for the improvement of the form-factor.

In addition, the Coulomb singularity gives rise to terms odd in $v$
starting at $\mathcal{O}(v^{-1})$, and the integrand must be
subtracted in a suitable way so that the numerical integral along the
contour that passes close to the singularity can be done accurately.
None of these difficulties occur in the matching calculations for the
space-like weak and electromagnetic form factors involving heavy
quarks.  We choose to match the real part of a suitable linear
combination of the electromagnetic form-factors $F_1(q^2), F_2(q^2)$ 
for time-like $q$ with $q^2 = 4M^2(1+v^2)$.

Earlier matchings of the vector annihilation current avoided these
issues by either being restricted to tree level
\cite{Bodwin:2002hg}
or to $v=0$
\cite{Jones:1998ub,Bodwin:ups}.
Neither is satisfactory: $v^2$ and $\alpha_s$ are comparable at around
$0.1$ in the $\Upsilon$ system and failure to include both leads to
strong discretisation errors in calculations of the leptonic width
\cite{Gray:2005ad}.
In addition, in \cite{Bodwin:ups} only the simplest NRQCD action was
used, keeping only terms to leading order in $v$.

This study corrects this, using a gauge action that allows lattice
matrix elements to be calculated using the state-of-the-art lattice
QCD ensembles produced by the MILC collaboration. The NRQCD action is
the same improved form used in recent studies, e.g.
\cite{Gulez:2003uf,Gray:2005ad}.
When the matching coefficients calculated here are married to lattice
NRQCD matrix elements, it will allow a determination of the leptonic
width that is correct to $\mathcal{O}(10\%)$ and of the ratio of the
widths of the 2S and 1S $\Upsilon$ states correct to within a few per
cent. The size of these uncertainties matches those in the
experimental measurements
\cite{Gray:2005ad},
which justifies our one-loop, perturbative approach in the matching.

The structure of the paper is as follows. In Sec.~\ref{sec_matching}
we describe the matching procedure. The continuum QCD matrix element
analytic calculation is given in Sec.~\ref{sec_qcd}. In
Sec.~\ref{sec_nrqcd} we present the numerical calculation of the
corresponding NRQCD matrix elements. The final matching coefficients
are determined in Sec.~\ref{sec_results}, and discussed in
Sec.~\ref{sec_summary}. In the appendices, we describe the tests we
have applied in our calculation to ensure the correctness of the
Feynman rules and of the loop integration, and also to establish the
independence of the results on the gauge fixing and infrared
regulator.

A preliminary version of this work was presented in
Ref.~\cite{Hart:2006uj}.

\section{Matching S-Wave Decays}
\label{sec_matching}

The leptonic width of a heavy quarkonium state of mass $M_{\bar{Q}Q}$
is given in terms of a QCD matrix element $M_\text{ME}$ by
\be
\Gamma_{ee} = \frac{16\pi}{6 M_{\bar{Q}Q}^2} \left| M_\text{ME} 
\right|^2 e_Q^2 \alpha_\text{em}^2
\ee
where $e_Q$ is the electric charge of the heavy quark and
$\alpha_\text{em}$ the fine structure constant. The matrix element
represents the probability of the heavy quarks meeting and
annihilating, and in the simplest picture is represented by a
hydrogenic ``wavefunction at the origin'': $|M_\text{ME}|^2 \simeq
\psi^\dagger \psi(0)$.

To compare with the experimentally measured widths, we want to
calculate this matrix element in continuum QCD, in a way that embodies
all the non-perturbative dynamics. As explained in the introduction,
we cannot do this directly and must instead use lattice NRQCD
simulations.

The problem is that we do not know a priori which NRQCD current we
should use. Instead, we should consider a set of suitable currents
and separately calculate the matrix elements of each one using the
Monte Carlo generated ensemble. The true QCD matrix element is a
linear combination of these, and this paper provides the necessary
coefficients.

We choose our NRQCD currents to be
\be
\bm{J}_i = \bm{\sigma} \left( \frac{\Delta^2}{M^2} \right)^i
\label{eqn_nrqcd_currs}
\ee
where bold face symbols denote spatial 3-vectors and $M$ is the heavy
quark mass.

To convert our non-perturbative lattice NRQCD current matrix elements
into the corresponding QCD value, $M_\text{ME} = \matel{J^\qcd}$, we
need matching coefficients $a_i$ such that
\be
\matel{\bm{J}^\qcd} = \sum_i a_i \matel{\bm{J}_i} \; .
\label{eqn:nrexp1}
\ee
In this paper we fix them. 

When we calculate the matrix elements of $J_i$ in the simulation, the
mass $M$ will of course be replaced by a number. We may choose it to
be the bare mass or (less usually) the renormalised value. The $a_i$
will differ accordingly, so we will give separate results for the bare
and renormalised cases.

Our matching method is summarised as follows: the NRQCD matrix
elements each depend differently on the heavy quark velocity (at tree
level, for instance, $\matel{J_i} \propto v^{2i}$). By choosing the
$a_i$ appropriately we can match the QCD velocity dependence order by
order in $v^2$. We make this choice perturbatively, performing the
velocity matching at each order in $\alpha_s$ in turn
\cite{braaten:radiative}.

We start by expanding the currents and matching coefficients
as power series in $\alpha_s$:
\bea
a_i & = & \sum_n \alpha_s^n a_i^{(n)} \; ,
\nonumber \\
\matel{\bm{J}} & = & \sum_n \alpha_s^n \matel{\bm{J}}^{(n)} \; .
\label{eqn:pertexp1}
\eea
The superscript $(n)$ denotes the $\mathcal{O}(\alpha_s^n)$
perturbative contribution. 

Working in the Breit frame (where the decaying meson is stationary),
we take the Euclidean four momentum of the quark and antiquark to be
\be
p^\mu = \left( iE_0,\pm \bm{p} \right) \; , \quad
\bm{p} = (0,0,aMv) \; .
\label{eqn_momentum}
\ee
The dimensionless expansion parameter is $v$. Although we refer to it
as the heavy quark velocity, it is related to the true velocity by
$v=\beta/\sqrt{1-\beta^2}$.

We treat the heavy quarks as being on shell. This is exact, even
though we might expect off shell contributions at
$\mathcal{O}(\alpha_s^2)$. By using the equations of motion, the
contributions from these within a bound state are seen to vanish at
all but a subset of spacetime points of measure zero
\cite{Lepage:private,Weisz:private}.

\subsection{Matching at tree level}

The matching at tree level is essentially trivial.  Using the standard
Dirac representation for the $\gamma$-matrices in terms of Pauli
$\sigma$-matrices:
\be
\gamma^0 = \left( 
  \begin{array}{rr}
    1 & 0 \\
    0 & -1
  \end{array}
\right) \; , \quad 
\gamma^i = \left(
  \begin{array}{rr}
    0 & \sigma^i \\
    -\sigma^i & 0
  \end{array}
\right) \; ,
\ee
the spinors become
\bea
u(\mathbf{p}) & = & \left(\begin{array}{c} \psi \\
\Di\frac{{\bm\sigma}\cdot\mathbf{p}}{E+M}\psi\end{array}\right)
\sqrt{\frac{E+M}{2E}} \; ,
\nonumber \\
v(\mathbf{p}) & = & \left(\begin{array}{c}
\Di\frac{{\bm\sigma}\cdot\mathbf{p}}{E+M}\chi \\
  \chi \end{array}\right) 
\sqrt{\frac{E+M}{2E}} \; ,
\eea
where $\psi$ and $\chi$ are the standard Pauli spinors for
quarks and antiquarks, respectively. We have chosen the
non-relativistic normalisation for consistency with NRQCD, since the
Foldy-Wouthuysen-Tani transformation
\cite{Pryce:1948pf,Tani:1949,Tani:1951,Foldy:1949wa} 
is unitary.

In terms of these Pauli spinors, the relevant Dirac tensor components
of the non-relativistic expansion of the tree-level matrix element
$\matel{J^{\qcd;\mu}}^{(0)} \equiv \bar{v}(\bm{-p}) \gamma^{\mu}
u(\bm{p})$ are:
\bea
\bar{v}(-\bm{p}) \gamma^{0} u(\bm{p}) & = & 0 \; ,
\nonumber \\
\bar{v}(-\bm{p}) \bm{\gamma} u(\bm{p}) & = &
\chi^\dagger \bm{\sigma} \left(\frac{2}{3}+\frac{M}{3E}\right)\psi
\nonumber \\
& \equiv & f_1(v^2) \; \chi^\dagger \bm{\sigma} \psi \; ,
\nonumber \\
\bar{v}(-\bm{p}) \frac{i\sigma^{i0}E}{M} u(\bm{p}) & = &
\chi^\dagger \sigma^i \left(\frac{E}{3M}+\frac{2}{3}\right) \psi
\nonumber \\
& \equiv & f_2(v^2) \; \chi^\dagger \sigma^i \psi \; .
\eea
where we have averaged over spatial directions for S-wave decays
\cite{Bodwin:2002hg}.

The tree-level matching coefficients must satisfy the
leading order term in Eq.~(\ref{eqn:nrexp1}):
\be
\matel{\bm{J}^\qcd}^{(0)} = \sum_i a_i^{(0)} \matel{\bm{J}_i}^{(0)}
\ee
The expansions in powers of $v$ are
\bea
f_1(v^2) & = & 1 - \frac{1}{6} v^2 + \frac{1}{8} v^4 + \mathcal{O}(v^6) \; ,
\nonumber \\
f_2(v^2) & = & 1 + \frac{1}{6} v^2 - \frac{1}{24} v^4 + \mathcal{O}(v^6) \; .
\eea
Using Eq.~(\ref{eqn_momentum}), the tree level NRQCD matrix elements
can be written as
\be 
\matel{\bm{J}_i}^{(0)} = g_i(v) \; \chi^\dagger \bm{\sigma} \psi \; .
\label{eqn_tree_level}
\ee
The tree level velocity dependence is
\bea
g_0(v) & = & 1 
\nonumber \\
g_1(v) & = & -\frac{4}{(aM)^2} \sin^2 \left(\frac{aMv}{2}\right)
\nonumber \\
g_2(v) & = & \frac{4}{(aM)^4} \left[ 4 \sin^2 \left( \frac{aMv}{2} \right) 
- \sin^2 \left( aMv \right) \right]
\eea
such that $g_i(v) = (-v^2)^i$ at lowest order in $v$.

A term by term comparison of these expansions with that of $f_1$ yields
\be
a_0^{(0)} = 1 \; , \quad
a_1^{(0)} = \frac{1}{6} \; , \quad
a_2^{(0)} = \frac{1}{8} - \frac{\left( aM \right)^2}{72} \; .
\ee

\subsection{Matching at one-loop order}

\begin{figure}[b]
\begin{center}
\includegraphics[width=\myfigwidth,clip]{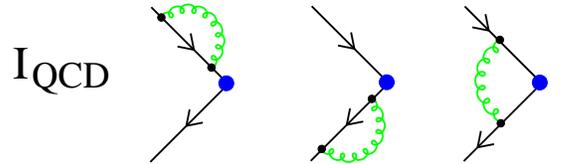}
\caption{One-loop corrections to the quark-antiquark annihilation
  current in QCD}
\label{fig:qcdann}
\end{center}
\end{figure}

To match at one-loop order, we need to calculate the one-loop QCD and NRQCD
corrections to the quark-antiquark annihilation vertex. The QCD corrections
consist of both self-energy insertions on the external legs and a vertex
correction, and for the case of a quark-antiquark vertex can be written as
\begin{widetext}
\bea
\matel{\bm{J}^\qcd}^{(1)} & = &
F^{(1)}_1(4E^2)  \; 
\bar{v}(-\bm{p}) \bm{\gamma} u(\bm{p})
+ i F^{(1)}_2(4E^2) \;
\bar{v}(-\bm{p}) \tilde{\bm{q}} u(\bm{p})
\nonumber\\
& = & \left[ F^{(1)}_1(4E^2) f_1(v^2) + F^{(1)}_2(4E^2) f_2(v^2) \right]
\chi^\dagger \bm{\sigma} \psi 
\label{eqn:qcddecomp1}
\eea
where $\tilde{q}_i\equiv \sigma_{i\nu}q^{\nu}/M$ and 
$F^{(1)}_{1,2}$ are the $\mathcal{O}(\alpha_s)$
contributions to the vertex structure functions.  We note that while
after renormalisation $F^{(1)}_1$ is UV-finite because of the Ward
identity, it contains IR divergences. These infrared divergences,
however, are the same as those that arise in NRQCD, since the
low-energy behaviour of the two theories is the same.

The $\mathcal{O}(\alpha_s)$ matching condition from
Eq.~(\ref{eqn:nrexp1}) is then
\bea
\sum_i a_i^{(1)} \matel{\bm{J}_i}^{(0)} & = &
\left[ F^{(1)}_1(4E^2) f_1(v^2) + F^{(1)}_2(4E^2) f_2(v^2) \right]
\chi^\dagger \bm{\sigma} \psi
- \sum_i a_i^{(0)} \matel{\bm{J}_i}^{(1)}
\nonumber \\
& \equiv & \left( \Iqcd - \Inrqcd \right) \chi^\dagger \bm{\sigma} \psi
\label{eqn:fitcoeff}
\eea
\end{widetext}
The infrared divergences cancel between the first two terms, leaving
an IR- and UV-finite expression that can be evaluated numerically.  We
opt to project out the $\sigma_2$ component and use the tree-level
expectation values of the NRQCD operators $J_i$ as our basis functions
to fit the difference between the QCD and NRQCD one-loop results and
determine $a_i^{(1)}$:
\be
\sum_i a_i^{(1)} g_i(v) = \Iqcd - \Inrqcd \; .
\label{eqn_fits}
\ee
To match to $\mathcal{O}(v^2)$ in this calculation $i$ runs from 0 to~1 only.

\section{Continuum QCD Calculation}
\label{sec_qcd}

To evaluate $\Iqcd$ analytically, we must regulate the infrared
Coulomb divergence in the Feynman integrals.  To avoid the
complications of twisted boundary conditions, we introduce a gluon
mass $\mu$ and use the gauge invariant St\"uckelberg propagator for
the massive vector field (see Sec.~(3-2-3) of
Ref.~\cite{Itzykson:book}):
\be
G_{\mu \nu} = 
\frac{g_{\mu \nu} - k_\mu k_\nu / \mu^2}{k^2 - \mu^2 + i \varepsilon}
+ \frac{k_\mu k_\nu / \mu^2}{k^2 - \mu^2/\lambda + i \varepsilon}
\ee
where $\lambda$ is the gauge fixing
parameter.

The one-loop QCD contribution is given by the sum of the Feynman
diagrams shown in Fig.~\ref{fig:qcdann}. The two leftmost rescale the
tree-level element by the quark wave function renormalisation constant
$Z$. The rightmost diagram is the one-loop vertex correction.  The
full one-loop vertex function is a rather formidable-looking
expression
\cite{karplus:fourthorder}. 
We know from the Ward identity, however, that the vertex function must
take the form
\be
\bar{u}(\bm{p'})\Gamma_{\mu}u(\bm{p})=
\bar{u}(\bm{p'})\left[F_1(q^2)\gamma_\mu+
\frac{i}{2M}F_2(q^2)\sigma_{\mu\nu}q^\nu\right]u(\bm{p})
\ee
when sandwiched between on-shell spinors, where $q=p-p'$ is the gluon
momentum flowing out of the vertex. We also know from the Ward
identity that $Z^{-1}=F_1(0)$, so that we can renormalise the vertex
function order by order by subtracting from $F_1(q^2)$ its value at
zero gluon momentum to obtain the renormalised structure function
\be
F^{(n),R}_1(q^2)=F^{(n)}_1(q^2)-F^{(n)}_1(0) \; .
\label{eqn_defn_F1r}
\ee
This amounts to including the effects of the first two diagrams, with
which we will therefore no longer concern ourselves.

For the case of quark-antiquark annihilation, we then have
\begin{widetext}
\be
\matel{J_\mu^\qcd}^{(1)}=
\bar{v}(-\bm{p})\left[F^{(1),R}_1(4E^2)\gamma_\mu
+\frac{iE}{M}F^{(1)}_2(4E^2)\sigma_{\mu 0}\right]u(\bm{p})
\ee
or, in terms of Pauli spinors
\be 
\matel{J_i^\qcd}^{(1)}=
\chi^{\dag}\sigma_i\psi
\left[F^{(1),R}_1(4E^2)f_1(v^2)+F^{(1)}_2(4E^2)f_2(v^2)\right] \; .
\ee
\end{widetext}
To compute $F_1$ and $F_2$ without resorting to the Feynman or
Schwinger parameter representations (which are not available for
NRQCD because the denominators are not quadratic), we employ a number
of techniques. A discussion of these will be useful later.

Since the decomposition of the vertex function into form factors
stated above is only valid between on-shell spinors, we put it between
the appropriate on-shell projectors
\begin{widetext}
\be
(\slashchar{p}'+M)\Gamma_{\mu}(p',p)(\slashchar{p}+M)=
(\slashchar{p}'+M)\left[F_1(q^2)\gamma_\mu
+\frac{i}{2M}F_2(q^2)\sigma_{\mu\nu}q^\nu\right](\slashchar{p}+M)
\ee
\end{widetext}
where the appropriate on-shell momenta for an incoming quark-antiquark
pair are given by $p_\mu=(E,\bm{p})$ and $p'_\mu=(-E,\bm{p})$ with
$E=\sqrt{M^2+\bm{p}^2}$.

Contracting the above equation with either $(p+p')^{\mu}$ or
$\gamma^{\mu}$, and taking the trace of both sides, we obtain two
equations for $F_1$ and $F_2$:
\bea
A & \equiv & \frac{(p+p')^{\mu}}{2M}\Tr((\slashchar{p}'+M)\Gamma_{\mu}(p',p)
(\slashchar{p}+M))
\nonumber \\
& = & \frac{(p+p')^2}{2M^2}(4M^2 F_1(q^2)-q^2F_2(q^2)) \; ,
\nonumber \\
B & \equiv &  
\Tr(\gamma^{\mu}(\slashchar{p}'+M)\Gamma_{\mu}(p',p)(\slashchar{p}+M))
\nonumber \\
& = & 4(2M^2+q^2)F_1(q^2)-6q^2F_2(q^2) \; ,
\eea
with solutions
\bea
F_1(q^2)&=&\frac{1}{4(4M^2-q^2)}\left[\frac{12M^2}{(p+p')^2}A-B\right]
\; ,
\nonumber \\
F_2(q^2)&=&\frac{2M^2(2M^2+q^2)}{q^2(p+p')^2}
\nonumber \\
&& \times \left[\frac{A}{(p+p')^2}
-\frac{B}{2(2M^2+q^2)}\right] \; .
\eea
At the one-loop level, the vertex function is given by the integral
expression (in Feynman gauge)
\be
\Gamma^{(1)}_{\mu}(p',p)=4 \pi C_2\int\frac{d^4k}{(2\pi)^4}
\frac{\gamma_\rho(\slashchar{l}'+M)
\gamma_\mu(\slashchar{l}+M)\gamma^\rho}{(k^2-\mu^2)(l'^2-M^2)(l^2-M^2)}
\ee
where we have defined the loop momenta $l=k+p$ and $l'=k+p'$ and
introduced a gluon mass $\mu$ as an infrared regulator. After
performing the manipulations outlined above, this vertex function
leads to 
\begin{widetext}
\bea
F^{(1)}_1(q^2) & = & 4\pi C_2 \int \frac{d^4k}{(2\pi)^4}
\frac{4}{(p+p')^2 (k^2-\mu^2) (l'^2-M^2) (l^2-M^2)} 
\nonumber \\
& \times &
\left[ 2M^2 l \cdot l' + 2 M^2 (p+p') \cdot (l+l') - 
\frac{6M^2}{(p+p')^2} (p+p') \cdot l \; (p+p') \cdot l' \right.
\nonumber \\
&&
\left. + 2p \cdot l' p' \cdot l - M^4 - M^2 p\cdot p' \right] \; ,
\nonumber \\
F^{(1)}_2(q^2) & = & 4\pi C_2 \int \frac{d^4k}{(2\pi)^4}
\frac{2M^2(2M^2+q^2)}{q^2 (p+p')^2 (k^2-\mu^2) (l^2-M^2) (l'^2-M^2)} 
\nonumber \\
& \times &
\left[ 4 \left( (p+p') \cdot (l+l') + l \cdot
  l'-M^2 \right) - \frac{2}{(p+p')^2} (p+p') \cdot l \; (p+p') \cdot l' \right.
\nonumber \\
&& \left. -\frac{8}{2M^2+q^2} \left( M^2 l \cdot l' - 2p \cdot l' \; p'\cdot l 
+ M^2(p+p') \cdot (l+l')+p \cdot p' - 2M^2 \right) \right]
\label{eqn_qcd_integrands}
\eea
\end{widetext}
for the structure functions.

In the physical limit $\mu\to 0$, the one-loop structure functions are
of course well-known analytically, since they are just the QED
structure functions multiplied by the group-theoretic factor $C_2 = 4/3$:
\bea
F^{(1),R}_1(q^2) & = & \frac{g^2 C_2}{4\pi^2}
\left[ \left( \log \frac{\mu}{M} +
 1 \right)
(\theta\cot\theta - 1) \right.
\nonumber \\
&& \left. + 2 \cot\theta\int_0^{\theta/2}\phi\tan\phi \,d\phi +
 \frac{\theta}{4}\tan\frac{\theta}{2} \right]
\nonumber \\
F^{(1)}_2(q^2) & = & \frac{g^2C_2}{8\pi^2}\frac{\theta}{\sin\theta}
\eea
where
\be
\theta=2\arcsin(E/M)
\ee
We have compared our numerical evaluation of the structure functions
in both the form factor and annihilation channels with their
analytical values and have found excellent agreement. Especially, we
were able to replicate the infrared divergence by varying our gluon
mass $\mu$. Resolving the $1/v$ Coulomb singularity in the
annihilation channel requires special care. To avoid a contamination
of the low-$v$ behaviour by the gluon mass $\mu$, which acts as a
cut-off on the $v$ dependence by limiting the momentum of the
exchanged gluon, we have scaled $\mu$ with $v$, and then were able to
observe the correct Coulomb singularity behaviour in the infrared
finite part of $F_1$.

\subsection{Wick rotation}

\begin{figure*}[t]
\begin{center}
\includegraphics[width=2\myfigwidth,clip]{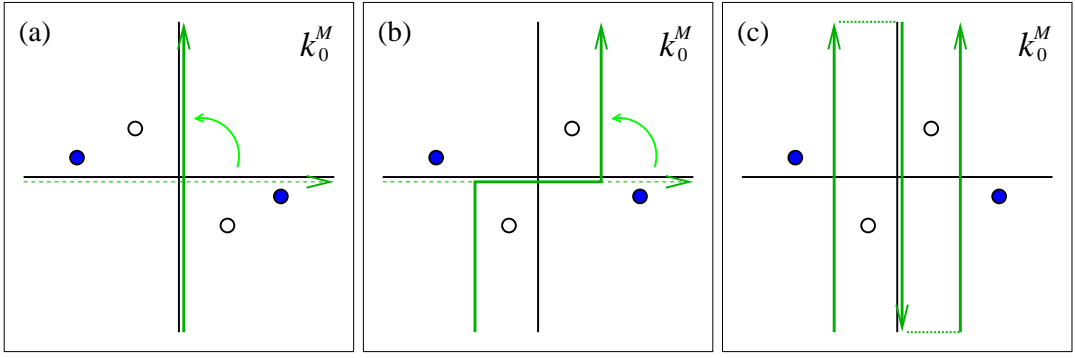}
\end{center}
\caption{\label{fig:qcdpoles} Locations of the poles and 
  choice of integration contour for the (Minkowski metric) $k_0$
  integration in QCD quark-antiquark annihilation. The solid circles
  represent poles in the gluon propagator, and the open circles the
  fermionic poles for various $|\bm{k}|$.}
\end{figure*}

In doing these calculations, we must be careful how we Wick rotate our
integration contour.  In the quark-antiquark annihilation channel, the
poles of the integrands in the complex $k_0$ plane are located as
shown in Fig.~\ref{fig:qcdpoles}. For $(\bm{k}+\bm{p})^2 > \bm{p}^2$,
the poles are all located second and fourth quadrants of the Argand
diagram for $k_0$, and the usual Wick rotation of the integration
contour is possible as in Fig.~\ref{fig:qcdpoles}a. When
$(\bm{k}+\bm{p})^2 \leq \bm{p}^2$, the fermionic poles cross the
imaginary $k_0$ axis and we need to be more careful and deform the
contour as per Fig.~\ref{fig:qcdpoles}b. This choice of contour is,
however, impractical. The short piece of the contour running along the
real axis is by far the most dominant contribution to the integral.
We will estimate the value of the integral using Monte Carlo methods.
To get this contribution correctly, we need to sample all
three-momenta along the contour with comparable weights. We therefore
use the equivalent contour shown in Fig.~\ref{fig:qcdpoles}c, which
works much more efficiently.

In this triple contour case, we choose the outlying contours to be
midway between the gluonic and fermionic poles. The St\"uckelberg
gluon propagator has two poles: one associated with the physical gluon
mass, and a second at $\mu^2/\lambda$. To avoid possible numerical
instabilities, we use the smaller of the two to fix the position of
the outer two contours.

Note that if we work with $v=0$ as in Ref.~\cite{Jones:1998ub} we can
always Wick rotate as per Fig.~\ref{fig:qcdpoles}a. For non-zero $v$,
however, it is important to note that the choice of an appropriate
contour is essential to obtain the correct result: with a naive
standard Wick-rotation, the structure functions obtained would not
even be Lorentz-invariant. We have explicitly checked that our choice
of contours leads to structure functions that are invariant under a
Lorentz boost. Another point to note is that even in the quark form
factor channel at spacelike $q^2$, where the quark poles do not cross
each other, a standard Wick rotation about the origin is not correct,
and the rotated contour has to be shifted along the real axis by an amount
depending on the kinematic frame, in order to pass between the poles and pick 
up the correct result.

\section{Lattice NRQCD Calculation}
\label{sec_nrqcd}

\begin{figure}[b]
\begin{center}
\includegraphics[width=\myfigwidth,clip]{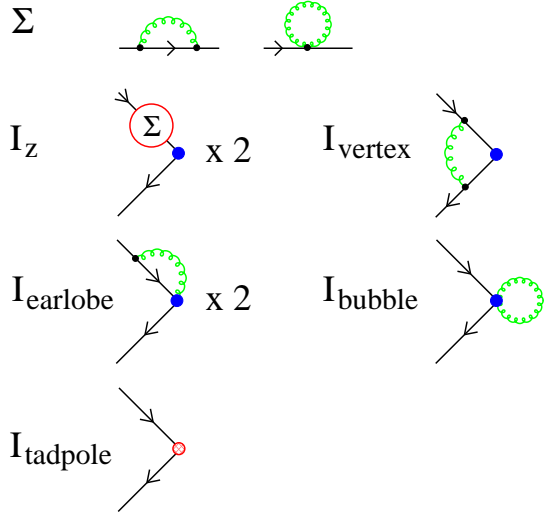}
\end{center}
\caption{\label{fig:nrqcdann} One-loop corrections to the self energy and 
  annihilation current in NRQCD. The gluons in these diagrams can be
  temporal as well as spatial. The solid (blue) circles represent the
  current in Eq.~(\ref{eqn:nrexp1}). The open (red) circle represents
  the contribution from tadpole improvement of the current. ``$\times
  2$'' denotes a similar diagram on the outgoing fermion line.}

\end{figure}

In this section we describe the perturbative calculation using the
lattice NRQCD action.

\subsubsection{The NRQCD Action}

The NRQCD action we consider is the same as Gulez et al. 
\cite{Gulez:2003uf},
and also the same as has been used in recent simulations
\cite{Wingate:2002fh,Gray:2005ad}
(although there is a typographical error in the description in the
latter
\cite{Davies:2006private}):
\bea
\mathcal{S}_\nrqcd & = & \sum_{x,t} \psi^\dagger \psi -
  \psi^\dagger
  \left( 1 - \frac{a\delta H}{2} \right) 
  \left( 1 - \frac{aH_0}{2n} \right)^n 
  \nonumber \\
  && \times \;
  U_4^\dagger 
  \left( 1 - \frac{aH_0}{2n} \right)^n
  \left( 1 - \frac{a \delta H}{2} \right)
  \psi \; ,
\label{eqn_nrqcd_action}
\eea
The $\psi^\dagger$ field is understood to be be located at
$(t,\bm{x})$, with the position of $\psi$ on the timeslice $t-1$ fixed
by gauge invariance. Other than consistency with previous work, there
are no strong arguments for the relative ordering of the kinetic and
interaction terms in the action. The ordering here differs from, for
instance, Ref.~\cite{Lepage:1992tx}.

The leading kinetic term is
\be
H_0=-\frac{\Delta^2}{2aM} 
\ee
where $M$ is the bare mass and $n$ is a stability parameter for the
nonrelativistic evolution equation, that must fulfil the condition
$n>3/(2aM)$ for the time reversal symmetric evolution equation
\cite{Thacker:1990bm,Lepage:1992tx}.
Gluonic corrections decrease the lower bound on $n$ to just above
$1/(aM)$
\cite{Davies:1991py}.

The time-reversal symmetric splitting of the $H_0$ operator either
side of the temporal link
\cite{Lepage:1992tx}
is designed to mimic the full time evolution due to $H_0$ along a
temporal lattice spacing in a way that avoids the well-known
instability in the discretisation of parabolic differential equations
(see, for instance, Sec.~19.2 of Ref.~\cite{press:numrec}). In this
way, the time step in the evolution equation is small enough to allow
the highest momentum modes in the theory to come into equilibrium,
whilst avoiding the need for a very small lattice spacing which makes
the theory too expensive to simulate.

The interaction term corrects for relativistic and discretisation
effects:
%
%
\newcommand{\delsq}{\Delta^{(2)}}
\newcommand{\delfour}{{\Delta^{(4)}}}
\newcommand{\Mbz}{{(aM)}}
\newcommand{\delv}{\bm{\nabla}}
\newcommand{\delvt}{\tilde{\bm{\nabla}}}
\newcommand{\Ev}{\tilde{\bm{E}}}
\newcommand{\Bv}{\tilde{\bm{B}}}
\newcommand{\sigmav}{\bm{\sigma}}
\begin{eqnarray}
a \delta H & = & 
- c_1 \, \frac{(\delsq)^2}{8\Mbz^3}
+ c_2\,\frac{i}{8\Mbz^2}\left(\delv\cdot\Ev - \Ev\cdot\delv\right) 
\nonumber \\
&&
 - c_3\,\frac{1}{8\Mbz^2} \sigmav\cdot(\delvt\times\Ev - \Ev\times\delvt)
\nonumber \\
&& 
- c_4\,\frac{1}{2\Mbz}\,\sigmav\cdot\Bv
\nonumber \\
&&
+ c_5\,\frac{\delfour}{24\Mbz}  - c_6\,\frac{(\delsq)^2}{16n\Mbz^2} \, .
\end{eqnarray}
We note that improved derivatives are used in the term proportional to
$c_3$ and that the improved field strength has not been rendered
explicitly traceless. The terms proportional to $c_i$ for $i=1...4$
provide relativistic corrections to $\mathcal{O}(Mv^4)$
\cite{Lepage:1992tx,Morningstar:1993de}
and represent the relativistic correction to the kinetic energy, the
nonabelian analogue of the Darwin term, the spin-dependent
interactions leading to spin-orbit couplings and the quark
chromomagnetic moment, respectively. The final two terms remove the
leading order discretisation error. All terms are understood to be
tadpole improved. We use the tree level values $c_i=1$, as in
Refs.~\cite{Gulez:2003uf,Gray:2005ad}. Other than tadpole improvement,
we do not consider the effects of radiatively correcting the $c_i$.

We obtain the Feynman rules for the NRQCD and gauge actions using an
automated procedure
\cite{Hart:2004bd},
as outlined in Appendix~\ref{sec_feyn_rules}. We also detail there the
tests we have carried out to ensure that the Feynman rule expressions
are correct, and the techniques we employ to speed up their evaluation
for specific momenta.

\begin{table*}[t]
\caption{\label{tab_diagrams} Diagrams contributing to 
matching calculation. Where no statistical error is given, 
it is smaller than the quoted precision of the number.}

\begin{ruledtabular}
  \begin{tabular}{lllllllllll}
    $aM$ & $v$ & $(\Iqcd - $ & $(\Ivertex - \Iin)$ & $\Iout$ &
    $\Iz$ & $\Iearlobe$ & $\Ibubble$ & $\Itadpole$ & $\Iqcd-\Inrqcd$ \\
    & & ~~~~$\Iodd)$ \\
\hline
4.0  & 0 & & & & & 0.02155~(1) & -0.06695~(3) & 0.04688 \\
     & 0.03 & -0.8511 & -1.3258~(28) & 1.2176~(5) & 1.8226~(9) & 0.0216 & -0.0668 & 0.0468 & -0.1315~(30) \\
     & 0.07 & -0.8611 & -1.3112~(39) & 1.2262~(5) & 1.8212~(9) & 0.0220 & -0.0661 & 0.0463 & -0.1452~(40) \\
     & 0.10 & -0.8738 & -1.2979~(44) & 1.2370~(5) & 1.8247~(9) & 0.0224 & -0.0652 & 0.0456 & -0.1615~(45) \\
     & 0.15 & -0.9044 & -1.2636~(55) & 1.2623~(5) & 1.8288~(9) & 0.0234 & -0.0631 & 0.0441 & -0.2004~(56) \\
\hline
2.8  & 0 & & & & & 0.05203~(2) & -0.13678~(5) & 0.09566 \\
     & 0.03 & -0.8511 & -1.3689~(21) & 0.9240~(4) & 1.6225~(9) & 0.0521 & -0.1365 & 0.0956 & -0.1736~(23) \\
     & 0.07 & -0.8611 & -1.3678~(29) & 0.9468~(4) & 1.6241~(9) & 0.0526 & -0.1359 & 0.0951 & -0.1809~(31) \\
     & 0.10 & -0.8738 & -1.3659~(33) & 0.9550~(4) & 1.6249~(8) & 0.0532 & -0.1349 & 0.0944 & -0.1862~(35) \\
     & 0.15 & -0.9044 & -1.3622~(39) & 0.9726~(4) & 1.6266~(9) & 0.0545 & -0.1327 & 0.0929 & -0.2007~(40) \\
\hline
1.95  & 0 & & & & & 0.12394~(3) & -0.28210~(6) & 0.19724 \\
     & 0.03 & -0.8511 & -1.3636~(15) & 0.7415~(2) & 1.3505~(8) & 0.1241 & -0.2820~(1) & 0.1971 & -0.1356~(17) \\
     & 0.07 & -0.8611 & -1.3669~(23) & 0.7459~(2) & 1.3486~(9) & 0.1246 & -0.2812~(1) & 0.1966 & -0.1357~(24) \\
     & 0.10 & -0.8738 & -1.3734~(26) & 0.7503~(2) & 1.3503~(9) & 0.1254 & -0.2804~(1) & 0.1960 & -0.1379~(27) \\
     & 0.15 & -0.9044 & -1.3979~(84) & 0.7634~(4) & 1.3537~(14) & 0.1271~(1) & -0.2780~(1) & 0.1944 & -0.1319~(86) \\
\hline
1.0  & 0 & & & & & 0.5093~(2) & -1.0720~(4) & 0.75000 \\
     & 0.07 & -0.8611 & -1.3681~(13) & 0.5004~(2) & 0.4335~(10) & 0.5103~(2) & -1.0713~(4) & 0.7494 & 0.4040~(17) \\
     & 0.10 & -0.8738 & -1.3956~(16) & 0.5034~(2) & 0.4323~(10) & 0.5110~(2) & -1.0701~(4) & 0.7488 & 0.4044~(19)\\
     & 0.15 & -0.9044 & -1.4146~(18) & 0.5127~(2) & 0.4323~(10) & 0.5132~(2) & -1.0676~(4) & 0.7472 & 0.4005~(21) \\
\end{tabular}
\end{ruledtabular}
\end{table*}
\subsubsection{The lattice gauge action}

To maintain compatibility with the MILC collaboration simulations, we
use the Symanzik improved gauge action
\bea
S_G & = &
-\beta \sum_{\genfrac{}{}{0pt}{2}{x}{\mu<\nu}} \left( 
\frac{5}{3} P_{\mu \nu}(x) - \frac{1}{12} R_{\mu\mu\nu}(x)
-\frac{1}{12} R_{\mu\nu\nu}(x) \right) 
\nonumber \\
&&
+ \mathcal{O}(\alpha_s)\; ,
\label{eqn_gauge_action}
\eea
where $P$, $R$ are $1 \times 1$ and $2 \times 1$ Wilson loops
respectively. $\mathcal{O}(\alpha_s)$ denotes possible radiative and
tadpole improvement of the action. As discussed later, these terms
will not contribute at one-loop.

The inverse lattice St\"uckelberg propagator is
\be
\Gamma^{\mu\nu}(k) = V^{\mu\nu}(k) + (a\mu)^2 \delta^{\mu\nu} + 
\lambda \hat{k}^\mu \hat{k}^\nu \; ,
\ee
where the two-point function $V^{\mu \nu}$ depends on the action
chosen. Gauge invariance requires the gauge fixing term to be
constructed from lattice momentum vector $\hat{k}^\mu \equiv 2 \sin
(ak^\mu/2)$, so the Feynman gauge ($\lambda=1$) propagator is only
diagonal for the Wilson gauge action (for which the gluon two-point
function is $V^{\mu\nu} = \hat{k}_\rho \hat{k}^\rho \delta^{\mu \nu} -
\hat{k}^\mu \hat{k}^\nu$). Note that $\lambda = 1/\alpha$ in the
notation of Ref.~\cite{Drummond:2002yg}.  As we do not consider Landau
gauge here, the inverse propagator is directly invertible and we do
not need to use an intermediate gauge.

\subsubsection{Annihilation currents and radiative improvements}

We use lattice NRQCD annihilation currents that are the naive
discretisations of Eq.~(\ref{eqn_nrqcd_currs}):
\bea
\bm{J}_0 & = & \sum_{x} \chi^\dagger_x \bm{\sigma} \psi_x \; ,
\nonumber\\
\bm{J}_1 & = & \sum_{x;i=1}^{~3}
\chi^\dagger_x \frac{\bm{\sigma}}{(aM)^2} 
\nonumber\\
&& \times \left(  
U_{i}(x) \psi_{x+\hat{\imath}} + 
U^\dagger_{i}(x-\hat{\imath}) \psi_{x-\hat{\imath}} 
- 2 \psi_x \right)
\label{eqn_defn_j1}
\eea
and the links in $J_1$ are understood to be tadpole improved.
Removing the mean field, ``tadpole'' contributions improves the
convergence of lattice perturbation theory markedly
\cite{Lepage:1992xa}.
Operationally, this is done by dividing every gauge link $U$ in the
action by a factor of $u_0$. Common definitions for $u_0$ are that it
is the mean link in Landau gauge or the fourth root of the mean
plaquette. We use the former, expanding the link perturbatively as
$u_0 \equiv 1 - \alpha_s u_0^{(2)} + \ldots$ with $u_0^{(2)} = 0.750$
from Ref.~\cite{Nobes:2001tf} and as used Ref.~\cite{Gulez:2003uf}.

Tadpole improvement of the NRQCD action does not contribute to our
calculation, as the fermion wavefunction renormalisation has no
tadpole correction for the time reversal symmetric form of the NRQCD
action (the argument mirrors the mean field analysis in
Ref.~\cite{Lepage:1992tx}).  As discussed before, we do not consider
any further radiative improvements of the NRQCD action.

Tadpole and other radiative improvements of the gauge action also do
not contribute to the matching calculation. The leading order effect
of these is an $\mathcal{O}(\alpha_s)$ insertion in the gluon
propagator. As there are no external gluons in our calculation, such
insertions will only contribute at two loops and above.

The only effect of tadpole improvement comes from the current $J_1$,
and its contribution to $\Inrqcd$ can be easily calculated:
\bea
\Itadpole & = & \frac{2 u_0^{(2)} a_1^{(0)}}{(aM)^2} 
\sum_{i=1}^3 \cos p_i \; ,
\nonumber \\
& = & -u_0^{(2)} a_1^{(0)} \left( 
-\frac{6}{(aM)^2} + v^2 + \mathcal{O}(v^4) 
\right) \; .
\eea
The only other possible source of radiative corrections comes from the
mass used in Eq.~(\ref{eqn_nrqcd_currs}) when we calculate the
non-perturbative NRQCD matrix elements in the Monte Carlo lattice
simulation. If the number $M$ used in the simulation is the
renormalised heavy quark mass, there is no further correction. If the
number for the bare mass is used instead, the renormalised mass is
$Z_M M$ and we should divide the matching coefficient $a_i$ by
$(Z_M)^{2i}$.  In this study, that amounts to shifting $a_1^{(1)}
\rightarrow a_1^{(1)} - 2 a_1^{(0)} Z_M^{(1)}$. We calculate the
multiplicative mass renormalisation factor in
Appendix~\ref{sec_renorm_prop}, and will present our results for the
matching coefficients both with and without the shift.

\subsection{Calculating the vertex corrections}

The one-loop diagrams contributing to the quark-antiquark annihilation
amplitude in NRQCD are shown in Fig.~\ref{fig:nrqcdann}. The NRQCD
$k_0$ integrals are around the unit circle in the $e^{ik_0a}$-plane.
The quark poles sometimes cross the unit circle (just as the they
crossed the imaginary axis in the QCD integrals), so we scale the
circle of integration to avoid them and adopt a similar triple-contour
strategy: integrating along three appropriately scaled concentric
circles when the poles cross each other, and along the unit circle
otherwise.

The $g_i(v)$ are all even functions of $v$, but $\Iqcd$ and $\Inrqcd$
both contain odd powers. We must assure ourselves that these exactly
cancel in Eq.~(\ref{eqn:fitcoeff}). The argument is that NRQCD is an
effective theory of QCD which can be systematically improved to
reproduce all features of QCD, including the odd powers. There are,
however, no S-wave operators containing odd powers of $v$ that we
could use in the improvement. The odd powers must therefore cancel
exactly in Eq.~(\ref{eqn:fitcoeff}). This is not entirely surprising
given that the odd powers arise from an even polynomial in $v$
multiplied by the $1/v$ Coulomb IR divergence, and we know that NRQCD
must reproduce the IR physics exactly. Nonetheless, it is worth
examining the cancellation in more detail.

Consider the power-expansion of the QCD expression 
\be
F_1^{(1),R}(4E^2)f_1(v^2) + F_2^{(1)}(4E^2)f_2(v^2) 
\ee
with $F_1^{(1),R}$ defined in Eq.~(\ref{eqn_defn_F1r}).  Using the
analytic results given above, we see that both $f_1(v^2)$ and
$f_2(v^2)$ contain only even powers of $v$. Any odd powers in the
expansion must therefore come from $F_1^{(1),R}(4E^2)$ or
$F_2^{(1)}(4E^2)$.  The analytical evaluation of the structure
functions shows us that $F_2^{(1)}(4E^2)$ contains odd powers in $v$
only in its imaginary part, so the odd powers in the final answer must
come from $F_1^{(1),R}(4E^2)$.

On the NRQCD side, we know that any odd powers in $v$ must come from
the quark pole giving rise to the Coulomb singularity, since the
residues in the $k_0$-plane of all other poles can be expanded in
powers of $v^2$. The odd powers therefore originate exclusively from
integrals of the form 
\be 
\int \frac{d^3k}{(2\pi)^3}\frac{({\bm{k}}^2)^\alpha({\bm{p}}^2)^\beta}
{{\bm{k}}^2({\bm{k}}^2+2\bm{k}\cdot \bm{p}+i\epsilon)} \; .  
\ee
A careful analysis of these shows that only those
integrals with $\alpha=0$ contribute to the real part, whereas the
others (which are UV-divergent in the continuum) contribute only to
the imaginary part.  Since
\be 
\int \frac{d^3k}{(2\pi)^3}\frac{1}{{\bm{k}}^2({\bm{k}}^2+2\bm{k}\cdot
    \bm{p}+i\epsilon)} =\frac{1}{16|\bm{p}|} 
\ee
the only odd powers of $v$ in the NRQCD result will come from
multiplying powers of ${\bm{p}}^2$ in the numerator with the Coulomb
singularity. Expanding these, we find the same coefficients
multiplying each odd power of $v$ as in the above QCD result.

We note that to obtain correct results to order $v^{2n}$ we have to
use the correctly matched $\mathcal{O}(v^{2n})$ tree-level
annihilation operator. We must also use $\mathcal{O}(v^{2n})$
quark-gluon vertices in the diagram involving spatial gluons and the
$\mathcal{O}(v^{2n+2})$ quark propagator (the expansion of the latter
around the Coulomb singularity pole gives an $\mathcal{O}(v^{2n})$
contribution).

In summary, then, matching at tree-level to $\mathcal{O}(v^{2p})$
guarantees the cancellation of the odd powers at one-loop level to
$\mathcal{O}(v^{2p-1})$.

We will estimate the NRQCD loop integrals stochastically using the
adaptive Monte Carlo package called VEGAS
\cite{lepage:vegas,lepage:vegas2} 
(see Sec.~7.8 of Ref.~\cite{press:numrec} for further discussion).
These estimates of the NRQCD integrals will only converge if the
integrands are both finite and relatively smooth.  Both $\Iqcd$ and
$\Inrqcd$ have an infrared Coulomb divergence.  Although these are
formally regulated by the gluon mass, the integrands are still sharply
peaked, leading to unacceptably slow convergence of the numerical
integration.

As we have discussed, all odd powers of $v$ cancel pointwise in the
difference of the two integrands, $\Iqcd - \Inrqcd$, leaving a smooth
integrand. The obvious strategy is to numerically estimate the
difference as a single integral, remembering that the NRQCD integrand
is only defined inside the finite Brillouin zone.

Direct subtraction has problems. The NRQCD integrand is quite
complicated and time-consuming to evaluate for given momenta. This
limits the number of integration points that VEGAS can consider in a
set time. Conversely, the QCD integrand needs a large number of points
to accurately estimate the integral: the terms like $1/(p+p')^2$ in
Eq.~(\ref{eqn_qcd_integrands}) give rise both to an apparently UV
divergent contribution to the $1/v$ Coulomb singularity and $1/v^2$
term in the result. These terms, however, come with a factor of
$\cos\theta$ from the scalar products with $(p+p')$, and thus vanish
only after integration over all spatial angles.

If we directly subtract the integrands, we arrive at a function that
is both expensive to evaluate and needs many integration points to
converge.  To get round this, we use an analytic form of the QCD
structure functions and only evaluate the NRQCD integrals numerically.
For the latter we need to smooth out the regulated $1/v$ infrared
divergence by subtracting an integrand with the same low-momentum
structure. Fortunately, we can still cancel all the odd powers of $v$
from the NRQCD integrand by multiplying the integral to be subtracted
by an appropriate function of $v^2$:
\bea
\Iodd & = & 
\mathop{\mathbb{I}\mathrm{m}} \left\{
-\frac{4h(v^2)}{3}  \int \frac{d^4k}{(2\pi)^4}
\left( \bm{k}^2 + \mu^2 \right)^{-1}
\nonumber\right.\\
&& \times
\left( ik_0 - \frac{\bm{k}^2 + 2\bm{k} \cdot \bm{p}}{2M} \right)^{-1}
\nonumber\\
&& \times
\left.\left( ik_0 + \frac{\bm{k}^2 + 2\bm{k} \cdot \bm{p}}{2M} \right)^{-1}
\right\}\nonumber\\
& = & \frac{h(v^2)}{12v}
\eea
with
\be
h(v^2)=\frac{\left( 1 + 2v^2 \right)
\left( 1 + 2{\sqrt{1 + v^2}} \right) }{3(1 + v^2)}\;.
\ee
This is certainly sufficient for the low powers of $v^2$ in which we are interested
here.  By comparing the respective power-series expansions term by
term, it can easily be seen that the odd powers of $v$ are the same as
in the QCD result.

To evaluate Eq.~(\ref{eqn_fits}) we therefore take the difference of
$(\Iqcd - \Iodd)$ calculated analytically and $(\Inrqcd - \Iodd)$
estimated numerically.  Both expressions are even power series in $v$,
and the subtracted NRQCD integrand is now sufficiently smooth that no
change of variable in the momentum coordinate, designed to ``squash'' 
many evaluation points onto the contour in the neighbourhood of the pole
\cite{Luscher:1985wf},
is required. It is convenient to split $\Iodd$ into two integration
regions, within ($\Iin$) and outside ($\Iout$) the NRQCD Brillouin
zone $|k^\mu| \le \pi/(aM)$.

$\Ivertex$ and $\Iz$ separately have infrared ``cutting'' divergences
that cancel in their sum. Although the divergences are regulated by
the gluon mass, by evaluating $\Ivertex$ and $\Iz$ together we would
have a smoother integrand for VEGAS. We meet the same problem as
before, however: $\Ivertex$ has a relatively cheap integrand but the
VEGAS estimates are slow to converge. $\Iz$ converges quickly, but
taking derivatives of Feynman rules makes the integrand expensive to
evaluate. Therefore, we calculate the NRQCD integrals in
Fig.~\ref{fig:nrqcdann} separately using VEGAS, choosing the number of
integration points to give comparable statistical accuracy in the
results.

The final calculation is then made up of
\bea
\Iqcd - \Inrqcd & = & \left( \Iqcd - \Iodd \right) - 
\left( \Inrqcd - \Iodd \right) 
\nonumber \\
& = & \left( \Iqcd - \Iodd \right) - 
\nonumber \\
&& \left(
\left[ \Ivertex - \Iin \right] - \Iout + \Iz + \Iearlobe \right.
\nonumber \\
&& \left. + \Ibubble + \Itadpole \right) .
\eea
%

%
%
%
%
\begin{table}[t]
\caption{The matching coefficients, as a function of the 
\textit{renormalised} heavy quark mass, for the leptonic width ($a_i$) 
and leptonic width ratio ($b_i$). Note that 
$a_0^{0} = 1$, $a_1^{0} = b_1^{(0)} = \frac{1}{6}$, and that 
there is \textit{no} subtraction to prevent mixing down.}
\label{tab:match_renorm}
\begin{ruledtabular}
\begin{tabular}{ccccccc}
$Ma$ & $n$ & $a_0^{(1)}$ & $a_1^{(1)}$ & $b_1^{(1)}$ & $b_2^{(0)}$ \\
\hline
4.0  & 2 & -0.1288~(27) & -3.29~(29) & -3.27~(30) & -0.09722 \\
2.8  & 2 & -0.1732~(21) & -1.27~(21) & -1.24~(22) & ~0.01611 \\
1.95 & 2 & -0.1358~(16) & -0.02~(16) & ~0.00~(17) & ~0.07219 \\
1.0  & 4 & ~0.4056~(20) & -0.22~(16) & -0.29~(17) & ~0.11111 \\
\end{tabular}
\end{ruledtabular}
\end{table}

%
%
%
%
\begin{table}[t]
\caption{The matching coefficients, as a function of the 
\textit{bare} heavy quark mass, for the leptonic width ($a_i$) 
and leptonic width ratio ($b_i$). Note that 
$a_0^{0} = 1$, $a_1^{0} = b_1^{(0)} = \frac{1}{6}$, and that 
there is \textit{no} subtraction to prevent mixing down.}
\label{tab:match_bare}
\begin{ruledtabular}
\begin{tabular}{ccccccc}
$Ma$ & $n$ & $a_0^{(1)}$ & $a_1^{(1)}$ & $b_1^{(1)}$ & $b_2^{(0)}$ \\
\hline
4.0  & 2 & -0.1288~(27) & -3.32~(29) & -3.30~(30) & -0.09722 \\
2.8  & 2 & -0.1732~(21) & -1.35~(22) & -1.32~(22) & ~0.01611 \\
1.95 & 2 & -0.1358~(16) & -0.16~(16) & -0.14~(17) & ~0.07219 \\
1.0  & 4 & ~0.4056~(20) & -0.50~(16) & -0.56~(17) & ~0.11111 \\
\end{tabular}
\end{ruledtabular}
\end{table}

\section{Results}
\label{sec_results}

In this paper we present results for four choices of heavy quark mass:
$aM=4.0$, $2.8$, $1.95$ and $1.0$. The first three represent the
$b$-quark mass on the MILC improved staggered ensembles with $a \simeq
0.09$~fm (``fine''), 0.12~fm (``coarse'') and 0.17~fm
(``super-coarse'')
\cite{Gray:2005ad}.
Mass $aM=1$ represents the charm quark mass on the super-coarse
lattices.  In agreement with Ref.~\cite{Gray:2005ad}, we use $n=2$ for
all masses except $aM=1.0$, where $n=4$.

We choose IR gluon mass $(a \mu)^2=10^{-4}$ and use Feynman gauge
$\lambda=1$. In Appendix~\ref{sec_further_tests} we show that our
results do not depend on either of these choices. We also compare with
relevant existing results in the literature for $v=0$.

The NRQCD diagrams were evaluated for a range of velocities from
$v=0.03$ to $v=0.15$. In addition, we evaluated $\Ibubble$ and
$\Iearlobe$ at $v=0$. The results are shown in
Table~\ref{tab_diagrams}.  We extracted the matching parameters using
a linear fit as per Eq.~(\ref{eqn_fits}). The matching coefficients
$a_0^{(1)}$ and $a_1^{(1)}$ are given in Tables~\ref{tab:match_renorm}
and~\ref{tab:match_bare}. Results from the former are to be used when
the number for the renormalised heavy quark mass is used to construct
the currents in Eq.~(\ref{eqn_defn_j1}). Results from the latter are
to be used if the bare mass is instead employed.

The results in the renormalised mass case are shown graphically
in Fig.~\ref{fig_fit_tad}. We note that for smaller masses, the
coefficients of the $J_1$ current, $a_1^{(1)}$ and $b_1^{(1)}$, are
much smaller when renormalised masses are used.

We have checked that the fits are not biased by higher terms in the
velocity expansion. Note that whereas $-\Iz$, $-\Ivertex$ and
$-\Ibubble$ reduce monotonically as $aM$ is increased, $\Iout$,
$-\Iearlobe$ and $-\Itadpole$ grow.  Given that the result of combining
these will depend on $(aM)^2$, $(aM)^4$ and $1/(aM)^2$, it is not
surprising that the matching coefficients do not vary monotonically
with the heavy quark mass.

Our computations of these diagrams have been performed on the SunFire
Galaxy-class supercomputer at the Cambridge-Cranfield High Performance
Computing Facility using an implementation of the VEGAS algorithm
adapted to parallel computers using MPI (Message Passing Interface).

\subsection{Mixing downwards}

Whilst at tree level matrix elements of $J_1$ contribute only at
$\mathcal{O}(v^{2n})$, at higher loop orders there will be
contributions at lower orders of $v^2$. We call this ``mixing down''.
In the case of $J_1$, the integrals $\Iearlobe$, $\Ibubble$ and
$\Itadpole$ are only weakly momentum dependent and $\Ivertex$ also
makes a contribution at $v=0$.

This is theoretically inconvenient as we must redo all previous
calculations when we improve the current to higher orders of $v^2$ and
cannot easily compare the new numbers with the old to check for
consistency. We can get around this by introducing subtracted currents
to prevent this downward mixing of currents. Although not essential
for lattice Monte Carlo calculations, subtracted currents are also
useful here as they make the convergence of the double series in
$\alpha_s$ and $v^2$ in Eqn.~(\ref{eqn:nrexp1}) most explicit. Thus we
can expect that the matrix element of the subtracted $J_1$ will vary
as $v^2$ (to some order in $\alpha_s$).

We define the subtracted currents as $\bar{J}_i \equiv z_{ij} J_j$,
where the coefficients $z_{ij}$ are chosen to prevent this downward
mixing of currents at all radiative orders:
\be
\left| \matel{\bar{J}_i}^{(n)} \right| = v^{2i} + \mathcal{O}(v^{2(i+1)}) 
\quad \forall~n \; .
\ee
At tree level $z_{ij}^{(0)} = \delta_{ij}$. At higher loop level we
set $z_{ij}^{(n>0)} = 0$ for $j \ge i$, as we are only concerned with
preventing downward mixing. For $\mathcal{O}(v^2)$ matching, the
only non-trivial element is $z_{10}^{(1)}$, fixed by
\bea
& \left.
z_{1j}^{(0)} \matel{J_j}^{(1)} + z_{1j}^{(1)} \matel{J_j}^{(0)} 
\right|_{v=0} = 0 \; ,
\nonumber \\
& \Rightarrow
z_{10}^{(1)} = - \left. \matel{J_1}^{(1)} \right|_{v=0} \; .
\label{eqn_mixdown}
\eea
Note that in this calculation we consider only $\Ibubble$, $\Iearlobe$
$\Itadpole$ and $\Ivertex$, all at $v=0$. Data for $z_{10}^{(1)}$ are
given in Table~\ref{tab:mix_down}. The reader should note that the
numbers in column 3 are not exactly the sum of the numbers for $v=0$
in Table~\ref{tab_diagrams}. We have improved the accuracy of these by
extrapolating data for all $v$ to $v=0$ using $g_1(v)$.

Correcting for mixing down does not change the tree level matching
coefficients. The subtracted one loop factors are related to the
original numbers by
\bea
\bar{a}_0^{(1)} & = & a_0^{(1)} - z_{10}^{(1)} \; ,
\nonumber \\
\bar{a}_1^{(1)} & = & a_1^{(1)} \; .
\eea
As this
subtraction is less likely to be needed in a lattice evaluation of
NRQCD matrix elements, it has \textit{not} been applied to the results
in Tables~\ref{tab:match_renorm} and~\ref{tab:match_bare}.

%
%
%
%
\begin{table}[t]
\caption{The mixing down subtraction. All diagrams are evaluated at 
$v=0$. See the comment below Eqn.~(\ref{eqn_mixdown}) for details 
of column 3.}
\label{tab:mix_down}
\begin{ruledtabular}
\begin{tabular}{ccccc}
&& $\Ibubble + \Iearlobe +$ \\ 
$Ma$ & $n$ & $\Itadpole$ & $\Ivertex$ & 
$z_{10}^{(1)}$ \\
\hline
4.0  & 2 & 0.00146~(2) & -0.11889~(4)  & 0.11743~(5) \\
2.8  & 2 & 0.01094~(4) & -0.17265~(6)  & 0.16171~(8) \\
1.95 & 2 & 0.03907~(5) & -0.26196~(9)  & 0.22289~(11) \\
1.0  & 4 & 0.1870~(3)  & -0.82970~(26) & 0.6427~(4)\\
\end{tabular}
\end{ruledtabular}
\end{table}

\subsection{Matrix element ratios}

If we are only interested in the ratio of leptonic widths of, say,
$\Upsilon(2s)$ and $\Upsilon(1s)$, we do not care about the overall
normalisation of the matrix element (which is independent of the mass
of the meson). We can therefore express the ratio of leptonic widths
as a ratio of differently normalised matrix elements
\bea
\frac{M_\text{ME}}{a_0} & = & \la J_0 \ra + 
\frac{a_1}{a_0} \la J_1 \ra + 
\frac{a_2}{a_0} \la J_2 \ra
\nonumber \\
& \equiv & 
\la J_0 \ra + 
b_1 \la J_1 \ra + 
b_2 \la J_2 \ra \; .
\eea
The advantage of this is that for $\Upsilon$ states $v^2 \sim \alpha_s
\sim 0.1$. We can obtain a \textit{ratio} that is accurate to a few
per cent, $\mathcal{O}(1\%-5\%)$ (to two loops, effectively) by
knowing $a_0,a_1$ to one loop and $a_2$ to tree level. That is, by
knowing no more than we have already calculated in this paper:
\bea
b_1 & \equiv &
\frac{a_1}{a_0} = \frac{a_1^{(0)}}{a_0^{(0)}} +
\frac{\alpha_s}{a_0^{(0)}} \left[
a_1^{(1)} - \frac{a_1^{(0)} a_0^{(1)}}{a_0^{(0)}}
\right]  \; ,
\nonumber \\
b_2 & \equiv &
\frac{a_2}{a_0} = \frac{a_2^{(0)}}{a_0^{(0)}} \; .
\eea
We give these values for the unsubtracted currents in
Tables~\ref{tab:match_renorm} and~\ref{tab:match_bare}.  Note that the
inclusion of $J_2$ at this order does not affect $a_0^{(1)},
a_1^{(1)}$, as there is no mixing down at tree level.

%
%
%
%
\begin{figure}[b]
\includegraphics[width=\myfigwidth,clip]{full_extrap}
\caption{\label{fig_fit_tad}The fits to the tadpole improved data 
versus velocity dependence of $\matel{\bm{J}_1}^{(0)}$.}
\end{figure}

\begin{figure}[b]
\includegraphics[width=\myfigwidth,clip]{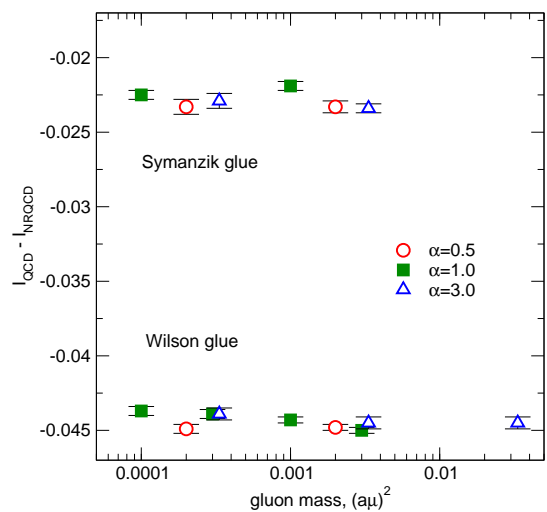}
\caption{\label{fig_jnw_total} $I_\qcd - I_\nrqcd$ 
for the NRQCD action described in Appendix~\ref{sec_gauge_coin} 
at $aM=2.1$ and $v=0.03$.} 
\end{figure}
\section{Summary and Conclusions}
\label{sec_summary}

In this paper we have presented a method to determine the QCD/NRQCD
matching coefficients for electromagnetic decays of heavy quarkonia in
lattice perturbation theory to order $\mathcal{O}(v^4,\alpha_s v^2)$.
This calculation was carried out for a realistic lattice NRQCD action
using largely automated methods for performing lattice perturbation
theory. 

The lattice NRQCD currents are given in Eq.~(\ref{eqn_defn_j1}). When
calculating their matrix elements in a lattice Monte Carlo simulation,
we have a choice as to whether we replace $M$ by the renormalised
heavy quark mass or the bare mass. If we choose $M$ to be the
renormalised heavy quark mass, the relevant matching coefficients are
given in Table~\ref{tab:match_renorm}. If the bare mass is used
instead, the matching coefficients include $Z_M$ and are given in
Table~\ref{tab:match_bare}.

We note that for the smaller quark masses, the $a_1^{(1)}$ and
$b_1^{(1)}$ coefficients of the current $J_1$ are very much smaller
when the renormalised quark mass is used.  This is particularly
relevant to NRQCD simulations of charm quarks on fine lattices, and
shows that the use of the renormalised rather than bare mass is a
major source of improvement in such simulations.

Individual Feynman diagrams vary monotonically with the mass, but when
combined together the competing dependencies lead to the final answer
varying as a complicated function of $M$.

We have performed a wide variety of checks of our calculation: we have
confirmed that the Feynman rules are correctly generated by comparing
with separately--obtained expressions in the literature, and that the
one-loop self energy renormalisation similarly agrees.  We have
checked that the infrared divergences vary as expected with changes in
the size of the regulating gluon mass and the choice of gauge. We have
also checked that the final answer is independent of both of these
factors. We have assured ourselves that the statistical errors quoted
by VEGAS are consistent with the size of variations in the Monte Carlo
estimates of the one-loop integrals.

These results could conceivably be checked using a series of
high-$\beta$ Monte Carlo simulations
\cite{Trottier:2001vj}.
Looking further, the computation of the perturbative one-loop
correction to the coefficient $c_1$ of the ${\bm\sigma}\cdot\bm{B}$
could be carried out using the methods employed in this paper.

\section*{Acknowledgments}

The authors thank G.P. Lepage and C.T.H. Davies for useful discussions
and the Cambridge--Cranfield High Performance Computing Facility for
advice and use of the SunFire supercomputer. A.H. thanks the U.K.
Royal Society for financial support. The University of Edinburgh is
supported in part by the Scottish Universities Physics Alliance
(SUPA). G.M.v.H. is supported in part by the Canadian Natural Sciences
and Engineering Research Council (NSERC) and by the Government of
Saskatchewan.

\appendix

\section{Feynman rules}
\label{sec_feyn_rules}

We use an automated method to obtain Feynman rules from the actions
and currents used in this calculation. The algorithm and its
implementation are described in Ref.~\cite{Hart:2004bd}. This allows
us to specify the action as a set of Wilson line contours that are
then Taylor expanded. The symmetries of the action are exploited to
produce very compact descriptions of the reduced vertex functions as
sums of $n$ monomials (each involving a relatively expensive
exponentiation).

The gluonic action expansion has been tested in a number of
calculations
\cite{Drummond:2002yg,Drummond:2002kp,Drummond:2003qu,Hart:2004jn}.
The expansion of the currents was checked by hand.

We tested the NRQCD action expansion by comparing with the Feynman
rules quoted in Eqs.~(A11-A36) of Ref.~\cite{Gulez:2003uf}. We find
complete agreement for general $c_i$, save in Eq.~(A33) which gives
the two gluon vertex for momenta specific to the gluon tadpole graph.
Our automated method shows this expression to be incomplete; it should
read:
\begin{widetext}
\bea
 \left[O_1 \right]^{(2)\mu ,\nu}_{s=0}(k,k,q,-q) &=&
\left( \frac{c_1}{2 (aM_0)^3} \, + \, \frac{c_6}{4n(aM_0)^2} \right)
\; 
\delta_{\mu,i} \, \delta_{\nu,j} 
\left[ \delta_{ij} \, \cos (k_j) 
\, \sum^3_{l=1}  \sin^2 (\frac{k_l}{2} )
 \; + \frac{1}{2}
\sin (k_i + \frac{q_i}{2}) \; \sin (k_j + \frac{q_j}{2}) \right]   
\nonumber \\
&+& \frac{ i c_2}{16 (aM_0)^2} \bigg[ (\delta_{\mu,j} \delta_{\nu,0} 
\, + \delta_{\mu,0} \delta_{\nu,j} ) 
\, \cos (k_j+\frac{q_j}{2}) \, \sin (q_j) \,
 \cos (\frac{q_0}{2}) \, \eta_{j0}   
\nonumber \\
&&  \qquad \qquad \qquad - \; \delta_{\mu,j} \delta_{\nu,j} \,
2 \, \cos (k_j+\frac{q_j}{2}) \, 
\sin (q_0) \, \cos (\frac{q_j}{2}) \, \eta_{j0}  \bigg]   
\nonumber \\
&+&  \frac{-c_5}{12 (aM_0)} \; \delta_{\mu,j} \delta_{\nu,j} \; 
 \left[ \cos (k_j) \, - \, \cos (2k_j) \,
 \cos ^2(\frac{q_j}{2}) 
+ \underline{\frac{1}{2} \sin \left( 2k_j \right) \sin \left(q_j \right)}
\; \right]
\eea
\end{widetext}
where the change is the addition of the final, underlined term. This
vanishes for $k=0$ and so does not affect the results in
Ref.~\cite{Gulez:2003uf}. Nonetheless, our detecting it highlights the
usefulness of an automatic action expansion program both for
developing new improved actions and for checking existing perturbative
results. We are happy to share copies of the program with interested
parties.

The NRQCD action in Eq.~(\ref{eqn_nrqcd_action}) naturally factorises
into the product of several distinct operators:
\be
S^\nrqcd = \sum_t \psi_t^\dagger\psi_t - 
\psi_t^\dagger \; A_tB_tU_4^\dagger B_{t-1}A_{t-1} \; \psi_{t-1} \; ,
\ee
where
\bea
A & = & \left(1-\frac{H_0}{2n}\right)^n \; ,
\nonumber \\
B & = & \left(1-\frac{a\delta H}{2}\right)  \; ,
\eea
and the subscript refers to the timeslice on which the fields are
located.

In the ``by-hand'' expansion it simplifies the algebra to derive
separate Feynman rules for $A$ and $B$ and combine them using the
convolution theorem
\cite{Morningstar:1993de,Gulez:2003uf}.
We also follow this approach: the $AB$ and $BA$ factors are on
different timeslices so no compression of the set of monomial factors
(``entities'') contributing to the reduced vertex function can occur.
Without such compression, it is computationally cheaper to calculate
the Feynman rules as a convolution of the expansions of $A$ and $B$.
The implementation of this has been checked by comparing with the
reduced vertex functions from the expansion of the full action.

Partial derivatives of the Feynman rules are computed automatically in
the code as per Ref.~\cite{Hart:2004bd}. We exploit the fact that the
velocity is purely along the $z$-axis to write
\be
\frac{\partial}{\partial \bm{p}^2} = \frac{1}{2p_3} 
\frac{\partial}{\partial p_3} \; .
\ee
The total on-shell derivative is implemented as
\bea
\frac{d}{d \bm{p}^2} & = & \frac{\partial}{\partial \bm{p}^2} + 
\frac{d p_0}{d \bm{p}^2} \frac{\partial}{\partial p_0} \; ,
\nonumber \\
\frac{d p_0}{d \bm{p}^2} & = &
\frac{i}{1 - T(\bm{p})} \frac{d T(\bm{p})}{d \bm{p}^2} \; ,
\eea
with $T(\bm{p}) = G_0^{-1}(0,\bm{p})$ coming from the bare fermion
propagator.

\section{Renormalising the fermion propagator}
\label{sec_renorm_prop}

In this appendix we review the one loop renormalisation of the fermion
propagator. The bare fermion propagator is
\be
aG_0^{-1}(p_0,\bm{p})=1 - z(1-aT(\bm{p}^2))
\label{eqn_bare_prop}
\ee
where $z = e^{-iap_0}$ and $T(\bm{p})$ is the kinetic energy.  The
$\mathcal{O}(\alpha_s)$ NRQCD quark self-energy can always be written
as
\bea
a\Sigma(p_0,\bm{p}) & = & A + B(p_0,\bm{p}) \; aT(\bm{p}) + 
\nonumber \\
&& C(p_0,\bm{p}) \left[ 1 - z \left(1 - aT(\bm{p}) \right) \right] \; ,
\eea
where $A$ is a constant. The resummed propagator is
\bea
aG^{-1}(p_0,\bm{p}) & = & aG_0^{-1}(p_0,\bm{p}) - 
\alpha_s a\Sigma(p_0,\bm{p})
\nonumber \\
& = & \left( 1 - \alpha_s (A+C) \right)
\left[ 1 - z \left( 1 + \alpha_s A \right) \right. \times
\nonumber \\
&& \quad \quad \quad \left.
\left( 1 - aT(\bm{p}) \left[ 1 - \alpha_s B/z \right] \right) \right] 
\nonumber \\
&& + \mathcal{O}(\alpha_s^2)
\label{eqn_sigma_prop}
\eea
In the infrared limit of small $\bm{p}^2$, this should be compared to
the renormalised form of Eq.~(\ref{eqn_bare_prop}):
\be
aG^{-1} = Z_\psi^{-1} \left( 1 - \bar{z} 
\left[ 1 - aT_R(\bm{p}) \right] \right) \; ,
\ee
with $T_R(\bm{p}) = \bm{p}^2 / (Z_M M)$.  Identifying $\bar{z} = z(1 +
\alpha_s A)$, the additive shift in the rest energy is
\bea
a\Delta E_\text{rest} & = & \ln (z/\bar{z})
\nonumber \\
& = & - \alpha_s A + \mathcal{O}(\alpha_s^2) \; ,
\eea
and $A = a\Sigma(p_0 = 0,\bm{p} = \bm{0})$.  The $p_0$ pole in the
propagator occurs at $\bar{z} = z_0 \equiv (1-aT_R)^{-1}$. The
wavefunction renormalisation is found by Taylor expanding
Eq.~(\ref{eqn_sigma_prop}) around this pole:
\bea
Z_\psi(\bm{p}) & = & 1 + \alpha_s \left( A + C + B \; aT - 
aT \bar{z} \frac{\partial B}{\partial \bar{z}} \right)_\text{on-shell}
\nonumber \\
& = & 1 + \alpha_s \left( 
a\Sigma + \frac{\partial a\Sigma}{\partial (iap_0)} \right)_\text{on-shell}
\label{eqn_Zpsi}
\eea
where the expressions are evaluated on the mass shell.  As the terms
in brackets are already $\mathcal{O}(\alpha_s)$, it is sufficient to
identify $T$ and $T_R$ and evaluate them at the pole of the bare
propagator $\bar{z} = (1-aT)^{-1}$. This result is general and
includes all orders in $v^2$ at $O(g^2)$. Morningstar
\cite{Morningstar:1993de} 
gives the expression for $Z_\psi$ at zeroth order in $v^2$ and our
result agrees with his to this order.

Working on the renormalised mass shell, the mass renormalisation
follows from
\be
\frac{1}{2 Z_M M} =
\left. \frac{d T_R}{d \bm{p}^2} \right|_{\bm{p^2}=0} 
= \frac{1}{2M} - \alpha_s 
\left. \frac{d (BT)}{d \bm{p}^2} \right|_{\bm{p^2}=0} \; .
\ee
We note that the total differential must also be evaluated on the
(bare) mass shell. From this we obtain
\be
Z_M = 1 + \alpha_s 2M \left. 
\frac{d a\Sigma}{d \bm{p}^2} \right|_{\bm{p^2}=0} \; .
\ee
Tadpole improvement affects $\Delta E_\text{rest}$ and $Z_M$, but not
$Z_\psi$ for NRQCD actions that are symmetric under time reversal
\cite{Lepage:1992tx}.
The one-loop contributions are given in Eqs.~(35,~36) of
Ref.~\cite{Gulez:2003uf}.

\section{Further code tests}
\label{sec_further_tests}

In this appendix we describe further, non-trivial tests of our
perturbative calculation that verify that our contour shifting and
numerical integration techniques are correct.

\subsection{One-loop self energy}
\label{sec_self_energy}

We have calculated the renormalisation of the NRQCD propagator as per
Appendix~\ref{sec_renorm_prop} using Feynman gauge and a gluon mass
of $(a\mu)^2 = 10^{-4}$.  The results are given in
Table~\ref{tab_qq_renorm}. For comparison, we also give the results of
Gulez et al.
\cite{Gulez:2003uf}.
Our data agree very closely. This provides further evidence that not
only are our Feynman rules correct, but also that we are combining
them correctly to form diagrams and evaluating the resulting integrals
correctly using VEGAS. We have also checked that the results are
correctly gauge variant and that the effect of the finite gluon mass
is negligible.

\begin{table*}[t]
\caption{\label{tab_qq_renorm} The renormalisation of the fermion 
propagator, as compared to Ref.~\cite{Gulez:2003uf}. 
Note that an IR factor corresponding to Eq.~(\ref{eqn_z_ir}) 
with $(a\mu)^2 = 10^{-4}$ has been applied to the subtracted 
data quoted in Ref.~\cite{Gulez:2003uf} for $Z_\psi^{(1)}(\bm{p}=0)$.}
\begin{ruledtabular}
\begin{tabular}{llllllll}
& & \multicolumn{2}{c}{$Z_\psi^{(1)}(\bm{p}=0)$} &
\multicolumn{2}{c}{$Z_M^{(1)}$} &
\multicolumn{2}{c}{$a\Delta E_0^{(1)}$} \\
$aM$ & $n$ & Us & Ref.~\cite{Gulez:2003uf} & 
Us & Ref.~\cite{Gulez:2003uf} & Us & Ref.~\cite{Gulez:2003uf} \\
\hline
4.0  & 2 & 1.8207 (7) & 1.813 (3) & 0.0817 (5)  & 0.082 (4) & 
0.8390 (1)  & 0.850 \\
2.8  & 2 & 1.6232 (6) & 1.617 (3) & 0.2350 (6)  & 0.235 (4) & 
0.7570 (10) & 0.767 \\
1.95 & 2 & 1.3494 (5) & 1.344 (3) & 0.4201 (8)  & 0.421 (4) & 
0.6765 (10) & 0.689 \\
1.0  & 4 & 0.4334 (7) & ---       & 0.8285 (16) & ---       & 
0.9684 (13) & ---   \\
     & 6 & ---        & 0.410 (3) & ---         & 0.859 (4) & 
---         & 0.758 \\
\end{tabular}
\end{ruledtabular}
\end{table*}

\subsection{Gauge covariance and invariance}
\label{sec_gauge_coin}

\begin{figure}[b]
\includegraphics[width=\myfigwidth,clip]{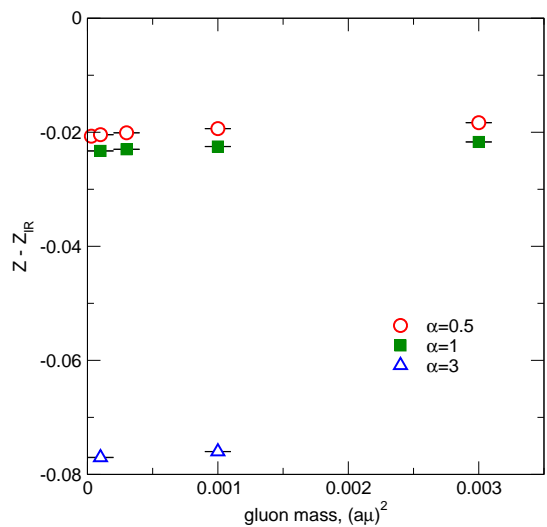}
\caption{\label{fig_jnw_lessir} $Z-Z_\text{IR}$
for the NRQCD action described in Appendix~\ref{sec_gauge_coin} 
at $aM=2.1$ and $v=0.03$ 
using the Wilson gauge action.} 
\end{figure}

\begin{figure}[b]
\includegraphics[width=\myfigwidth,clip]{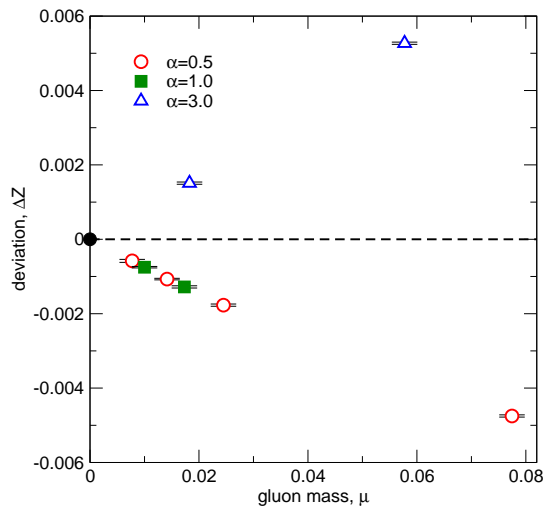}
\caption{\label{fig_jnw_deltaz} $\Delta Z$ 
for the NRQCD action described in Appendix~\ref{sec_gauge_coin} 
at $aM=2.1$ and $v=0.03$ 
using the Wilson gauge action.} 
\end{figure}

\begin{figure}[b]
\includegraphics[width=\myfigwidth,clip]{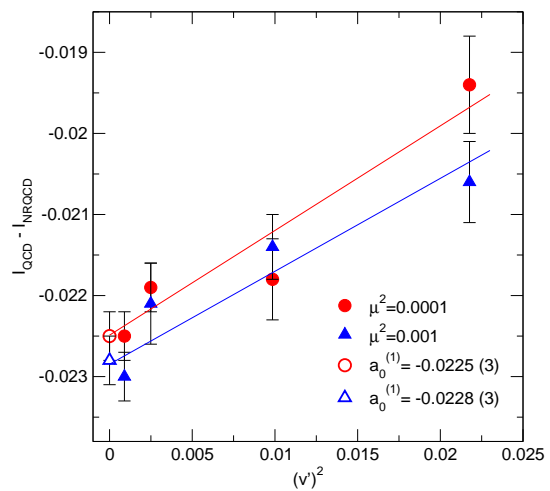}
\caption{\label{fig_jnw_extrap} Determination of $a_0^{(1)}$
for the NRQCD action described in Appendix~\ref{sec_gauge_coin} 
at $aM=2.1$ and $v=0.03$ 
using the Wilson gauge action.} 
\end{figure}

We have also looked closely at the effect of changing the gauge and
infrared regulator. For these tests, we use a simpler NRQCD action
with coefficients $c_i=0$ for $i=1 \ldots 4$ and $c_5 = c_6 = 1$, as
used in Ref.~\cite{Jones:1998ub}, with $n=2$ and $aM=2.1$. We used
both the Wilson and Symanzik-improved gauge actions and set current
$J_1 = 0$, which implies $\Iearlobe = \Ibubble = \Itadpole = 0$. We
use three choices of gauge: Feynman gauge $\lambda=1$, an unnamed
gauge with $\lambda=2$ and Yennie--Fried gauge $\lambda=\frac{1}{3}$.

Firstly, $I_\qcd - I_\nrqcd$ should be independent both of the choice
of gauge and the gluon mass. This is seen for $v=0.03$ in
Fig.~\ref{fig_jnw_total}. We note that the size of the scatter of
points about a single mean value is consistent with the statistical
errors assigned to the data points by VEGAS. This gives us some
confidence that these errors are not being underestimated in our
calculation.

Next, $Z$ and $V$ separately have infrared divergences that are
regulated by the gluon mass, but which cancel in $Z$+$V$. The
cancellation has already been shown by the absence of diverging
behaviour at small $a\mu$ in Fig.~\ref{fig_jnw_total}.

Here we check that the individual diagrams show the correct divergence.
We compare with the expectations for continuum QCD: lattice
NRQCD is an effective description of this and must preserve the
same infrared structure (up to possible discretisation errors of order
$a\mu$).  

Given the lack of overall divergence in $Z+V$, it is sufficient to
concentrate on $Z$, which is determined to greater statistical
accuracy.

The one-loop continuum expression is given in Eq.~(7-44) of
Ref.~\cite{Itzykson:book} 
(adding a colour factor of $4/3$):
\be
Z_\text{cont} = -\frac{1}{12\pi^2} \left(
\frac{1}{\lambda} \ln \frac{\Lambda^2}{M^2} + 3 \ln \frac{\mu^2}{M^2} - 
\frac{1}{\lambda} \ln \frac{\mu^2}{\lambda M^2} + \frac{9}{4} \right)
\ee
The infrared divergent contribution is
\be
Z_\text{IR} = -\frac{1}{12\pi^2} \left( 3 -
  \frac{1}{\lambda} \right) \ln \mu^2 \; , 
\label{eqn_z_ir}
\ee
which vanishes in Yennie--Fried gauge.

In Fig.~\ref{fig_jnw_lessir} we plot $Z$ with the continuum divergence
removed (replacing $\mu$ by $a\mu$). There is no discernible
divergence as $a\mu \to 0$. The slight gradient betrays a residual
dependence on the gluon mass. To emphasize this, we plot the deviation
$\Delta Z(a\mu) = Z(a\mu) - Z(10a\mu)$ in Fig.~\ref{fig_jnw_deltaz}.
The deviation disappears as we take $a\mu$ to zero and is a
discretisation effect.

\subsection{Current matching at $v=0$}

Finally, we have tried to verify the one-loop, $\mathcal{O}(v^0)$
annihilation current matching of Jones and Woloshyn
\cite{Jones:1998ub}.
Following the method in the main text, we get $a_0^{(1)} =
-0.0225~(3)$ for $(a\mu)^2=10^{-4}$ and $-0.0228~(3)$ for $(a\mu)^2 =
10^{-3}$ (using $n=2$ and $aM=2.1$ with the Symanzik gauge action).
The extrapolations to $v=0$ are shown in Fig.~\ref{fig_jnw_extrap}.
The statistical compatibility of the results shows that
$(a\mu)^2=10^{-4}$ is small enough that any residual gluon mass
dependence of the results is swamped by the statistical uncertainties
in the VEGAS integration. 

At the same parameter values, Jones and Woloshyn give
$a_0^{(1)}=-0.0253~(3)$ [inserting the appropriate number from
Table~II into their Eq.~(27)]. This result broadly agrees with ours,
which gives us confidence that there are no gross disagreements in our
methods. 

There is still a small, but apparently significant deviation, which we
also see at a second mass value.  The stringent tests described in
these Appendices were our attempt to account for this difference. As
already described, we have checked our Feynman rules are correct and
give the correct self energy (and derivatives). We find the correct
infrared divergences and Lorentz invariance. We have gauge invariance
and independence on the gluon mass regulator. We have also checked
that the statistical errors quoted by VEGAS are not underestimated.
In the light of these, we feel confident that our calculation is
correct.


\end{document}